\documentclass{article}
\usepackage{spconf}

\usepackage{color}
\usepackage{graphicx}
\usepackage{pifont}
\usepackage{lipsum}

\usepackage{array}
\usepackage{booktabs}
\usepackage{float}
\usepackage{footnote}
\usepackage{multirow}
\usepackage{makecell}
\makesavenoteenv{tabular}
\makesavenoteenv{table}
\usepackage[table]{xcolor}

\usepackage{amsmath,amsfonts,bm}

\usepackage{url}
\usepackage{hyperref}
\hypersetup{colorlinks=True, urlcolor=black}

\usepackage[ruled,vlined]{algorithm2e}
\usepackage{enumitem}
\def\ours{DS-WED}
\def\prosodyevalsamples{1000}
\def\prosodyevalratings{2000}

\title{Measuring Prosody Diversity in Zero-Shot TTS: A New Metric, Benchmark, and Exploration}

\name{
    Yifan Yang$^{1\star}$\thanks{$^\star$Equal Contribution\quad$^\dagger$Project Leader\quad$^\ddagger$Corresponding Author},
    Bing Han$^{1\star}$,
    Hui Wang$^{3}$,
    Long Zhou$^{2\dagger}$,
    Wei Wang$^{1}$,
    Mingyu Cui$^{2}$,
    Xu Tan$^{2}$,
    Xie Chen$^{1,4\ddagger}$\thanks{This work was done during internship at Tencent Hunyuan and was supported by the Science and Technology Innovation (STI) 2030-Major Project (2022ZD0208700), National Natural Science Foundation of China (No. U23B2018), Shanghai Municipal Science and Technology Major Project under Grant 2021SHZDZX0102, and Yangtze River Delta Science and Technology Innovation Community Joint Research Project (2024CSJGG1100).}
}

\address{
    $^{1}$X-LANCE Lab, MoE Key Lab of Artificial Intelligence, Jiangsu Key Lab of Language Computing,\\
    Shanghai Jiao Tong University\;
    $^{2}$Tencent Hunyuan\;
    $^{3}$Nankai University\;
    $^{4}$Shanghai Innovation Institute
}

\begin{document}
\ninept

\maketitle

\begin{abstract}
Prosody diversity is essential for achieving naturalness and expressiveness in zero-shot text-to-speech (TTS).
However, frequently used acoustic metrics capture only partial views of prosodic variation and correlate poorly with human perception, leaving the problem of reliably quantifying prosody diversity underexplored.
To bridge this gap, we introduce ProsodyEval, a prosody diversity assessment dataset that provides Prosody Mean Opinion Score (PMOS) alongside conventional acoustic metrics.
ProsodyEval comprises \prosodyevalsamples{} speech samples derived from 7 mainstream TTS systems, with \prosodyevalratings{} human ratings.
Building on this, we propose the Discretized Speech Weighted Edit Distance (\ours{}), a new objective diversity metric that quantifies prosodic variation via weighted edit distance over semantic tokens.
Experiments on ProsodyEval show that \ours{} achieves substantially higher correlation with human judgments than existing acoustic metrics, while remaining highly robust in speech tokenization from HuBERT and WavLM.
Leveraging \ours{}, we benchmark state-of-the-art open-source TTS systems on LibriSpeech test-clean and Seed-TTS test-en, and further explorations uncover several factors that influence prosody diversity, including generative modeling paradigms, duration control, and reinforcement learning. Moreover, we find that current large audio language models (LALMs) remain limited in capturing prosodic variations.
Audio samples are available at https://prosodyeval.github.io.
\end{abstract}

\section{Introduction}
Prosody plays a central role in spoken communication, conveying paralinguistic information such as emotion, attitude, and intent that shapes how listeners interpret meaning. Subtle variations in pitch, intensity, and duration can fundamentally alter the interpretation of an utterance even with identical text, making prosody crucial to naturalness and expressiveness in speech synthesis.
While zero-shot TTS~\cite{vallt, felle, istlm, streammel} has advanced in intelligibility and speaker similarity, prosody diversity has received far less attention.
Existing evaluation practices~\cite{prosodyevaluationreview} rely largely on log $F_0$ root mean squared error (RMSE), which correlates weakly with human judgments, captures pitch but neglects rhythm and intensity~\cite{phoneticsofprosody}, and requires costly dynamic time warping (DTW).
Consequently, a clear gap remains between automatic metrics and human perception of prosody diversity.

Recent efforts have shown that both continuous~\cite{superbprosody} and discrete~\cite{prosodydiscretespeech} self-supervised (SSL) speech representations~\cite{k2ssl} effectively encode prosodic information.
Concurrently, NLP automatic evaluation metrics have been adapted to discrete speech representations for reference-free~\cite{speechlmscore} and reference-aware~\cite{speechbertscore} speech quality assessment. In particular, SpeechTokenDistance~\cite{speechbertscore} is a simple approach but remains preliminary, as it focuses solely on measuring correlation with acoustic metrics without examining what the differences represent or further analyzing correlations with human perception.

With these perspectives in mind, we introduce ProsodyEval, a human-annotated dataset for prosody diversity assessment, and \ours{} (\underline{D}iscretized \underline{S}peech \underline{W}eighted \underline{E}dit \underline{D}istance), a new objective diversity metric that quantifies prosodic variation via weighted edit distance over semantic token sequences derived from speech SSL models such as HuBERT~\cite{hubert} and WavLM~\cite{wavlm} through $k$-kmeans clustering.
ProsodyEval contains \prosodyevalsamples{} synthetic samples from 7 mainstream TTS systems, paired with \prosodyevalratings{} human ratings of prosody diversity.
Experiments show that \ours{} achieves substantially higher correlation with human judgments than frequently used acoustic metrics, including log $F_0$ RMSE and Mel cepstral distortion (MCD), while remaining robust across models, layers, and cluster sizes.
Building on \ours{}, we establish the first systematic benchmark of prosody diversity across state-of-the-art open-source TTS systems on LibriSpeech \textit{test-clean} and Seed-TTS \textit{test-en}.
Our further explorations reveal the impact of modeling paradigms, duration control, and reinforcement learning (RL), and demonstrate that even advanced large audio language models (LALMs) like Gemini 2.5 Pro remain unreliable for prosody evaluation.

Our contributions\footnote{Code and models are publicly available at \href{https://github.com/yfyeung/DS-WED}{yfyeung/DS-WED}.} are threefold:
\begin{itemize}[leftmargin=*, itemsep=0pt, topsep=0pt]
\item We propose \ours{}, a reliable objective prosody diversity metric based on semantic token weighted edit distance. Through validation on ProsodyEval, a human-annotated dataset for prosody diversity in zero-shot TTS, we demonstrate \ours{} correlates better with human ratings than existing acoustic metrics.

\item We establish the first benchmark of prosody diversity across state-of-the-art open-source TTS systems, offering systematic comparisons and analyses of different generative paradigms.

\item We investigate key factors shaping prosody diversity, showing that (1) autoregressive (AR) systems outperform flow-matching based non-autoregressive (NAR), but not masked generative modeling (MGM), as duration variation is critical; however, NAR lacks duration modeling, while flow-matching models with implicit alignment suffer from inherent constraints; (2) RL via direct preference optimization (DPO) trades off diversity for intelligibility; and (3) current LALMs remain unreliable for prosody understanding.

\end{itemize}
\clearpage

\section{ProsodyEval Dataset}

\subsection{Data Collection}
We construct ProsodyEval dataset by aggregating synthetic speech from diverse generative paradigms, spanning AR and NAR approaches, including next-token prediction~\cite{xtts,cosyvoice,cosyvoice2}, flow matching~\cite{e2tts,f5tts,zipvoice}, and MGM~\cite{maskgct}. Concretely, ProsodyEval comprises samples synthesized by seven recent open-source TTS systems, namely XTTS-v2~\cite{xtts}, CosyVoice~\cite{cosyvoice}, CosyVoice 2~\cite{cosyvoice}, E2 TTS~\cite{e2tts}, F5-TTS~\cite{f5tts}, MaskGCT~\cite{maskgct}, and ZipVoice~\cite{zipvoice}. These paradigms reflect widely used approaches in modern TTS, rendering the dataset representative for evaluating prosodic variation.

Each system generates a group of five samples per input with random seeds from 0 to 4, using prompt speech and text from LibriSpeech \textit{test-clean}~\cite{librispeech} and Seed-TTS \textit{test-en}~\cite{seedtts}.  
To exclude synthesis errors, we filter all groups to retain only those in which every sample is subjectively perceived as word-by-word aligned with the text.
In total, ProsodyEval contains \prosodyevalsamples{} synthetic speech samples.

\subsection{PMOS Collection}
We recruit 20 graduate students with research experience in TTS as raters and conduct a MOS test focusing on prosodic differences. In total, \prosodyevalratings{} high-quality ratings are collected.

\noindent\textbf{Evaluation Dimension}\quad
The prosodic difference score measures the extent of variation in pitch, rhythm, and stress patterns between two audio samples generated with the same model, text, and speaker.
A score of 1 indicates nearly identical prosody with imperceptible variation, while a score of 5 reflects clear and consistent prosodic differences, suggesting noticeably distinct speaking styles.

\noindent\textbf{Listening Test Design}\quad
All MOS tests are conducted online under controlled conditions. After a short training session, all raters are instructed to perform the assessments in a quiet environment.
In each trial, a group of five audio samples generated by the same system and speaker is presented. Raters perform pairwise comparisons across all ten possible pairs within the group in random order. Each pair is rated on a five-point Likert scale according to the perceived prosodic difference.
To aid judgment, both audio playback and corresponding waveform visualization are provided.
Raters are encouraged to replay samples as needed, enabling them to refine their judgments and capture both subtle and pronounced differences across pairs.

\subsection{Acoustic Prosody Metrics}
ProsodyEval incorporates two frequently used acoustic metrics, namely log $F_0$ RMSE and MCD.
For each pair of samples within a group, FastDTW~\cite{fastdtw}, a linear-time approximation of DTW, is first applied to align their lengths, after which log $F_0$ RMSE and MCD are computed.  
Together, these two metrics serve as complementary baseline indicators for evaluating prosodic variation.
\newcommand{\setucell}[4]{%
  \expandafter\gdef\csname cell@#1@#2@color\endcsname{#3}%
  \expandafter\gdef\csname cell@#1@#2@text\endcsname{#4}%
}
\newcommand{\useucell}[2]{%
  \cellcolor{\csname cell@#1@#2@color\endcsname}{\csname cell@#1@#2@text\endcsname}%
}
\newcommand{\showcell}[2]{%
  \ifnum#1=#2 -%
  \else
    \ifnum#1<#2 \useucell{#1}{#2}%
    \else        \useucell{#2}{#1}%
    \fi
  \fi
}

\setucell{1}{2}{green!10}{0.30{\tiny [0.19,\;0.40]}}
\setucell{1}{3}{green!25}{0.66{\tiny [0.58,\;0.73]}}
\setucell{1}{4}{green!45}{\textbf{0.77}{\tiny [0.73,\;0.81]}}
\setucell{2}{3}{green!15}{0.35{\tiny [0.25,\;0.44]}}
\setucell{2}{4}{green!15}{0.36{\tiny [0.26,\;0.45]}}
\setucell{3}{4}{green!50}{0.82{\tiny [0.74,\;0.87]}}

\begin{table}[t]
\centering
\small
\caption{
Correlation matrix of average Pearson coefficients ($\bar{r}$) between human ratings and objective metrics, aggregated across groups via Fisher’s $Z$ transformation.
Values in brackets denote 95\% confidence intervals, computed in the Fisher space and back-transformed. 
Darker shades of green indicate stronger correlations.
All correlations are statistically significant at $p < 0.001$.
The strongest correlation with human ratings is highlighted in \textbf{bold}.
}
\vspace{2pt}
\renewcommand\tabcolsep{3pt}
\resizebox{\linewidth}{!}{
\begin{tabular}{lcccc}
\toprule[1pt]
 & PMOS & log $F_0$ RMSE & MCD & \ours{} \\
\midrule
PMOS             & \showcell{1}{1} & \showcell{1}{2} & \showcell{1}{3} & \showcell{1}{4} \\
log $F_0$ RMSE    & \showcell{2}{1} & \showcell{2}{2} & \showcell{2}{3} & \showcell{2}{4} \\
MCD               & \showcell{3}{1} & \showcell{3}{2} & \showcell{3}{3} & \showcell{3}{4} \\
\ours{}           & \showcell{4}{1} & \showcell{4}{2} & \showcell{4}{3} & \showcell{4}{4} \\
\bottomrule[1pt]
\end{tabular}
}
\vspace{-1.5em}
\label{tab:corr}
\end{table}
\begin{table*}[t]
\centering
\small
\caption{
Prosody diversity benchmark of zero-shot TTS systems from diverse generative paradigms on LibriSpeech \textit{test-clean} and Seed-TTS \textit{test-en}, using traditional acoustic metrics log $F_0$ RMSE and MCD as well as our proposed \ours{}, each computed as micro-average (\textit{Avg.}) and rank-based score (\textit{Borda Avg.}). The best results are highlighted in \textbf{bold}, and the second-best are \underline{underlined}.
}
\vspace{2pt}
\renewcommand\tabcolsep{2pt}
\resizebox{\linewidth}{!}{
\begin{tabular}{l*{12}{c}}
\toprule[1pt]
\multirow{3}{*}{\textbf{System}}
& \multicolumn{6}{c}{\textbf{LibriSpeech \textit{test-clean}}} & \multicolumn{6}{c}{\textbf{Seed-TTS \textit{test-en}}} \\
\cmidrule(lr){2-7}\cmidrule(l){8-13}
& \multicolumn{2}{c}{\textbf{log $F_0$ RMSE}}
& \multicolumn{2}{c}{\textbf{MCD}}
& \multicolumn{2}{c}{\textbf{\ours{}}}
& \multicolumn{2}{c}{\textbf{log $F_0$ RMSE}} 
& \multicolumn{2}{c}{\textbf{MCD}}
& \multicolumn{2}{c}{\textbf{\ours{}}} \\
\cmidrule(lr){2-3}\cmidrule(l){4-5}\cmidrule(lr){6-7}\cmidrule(l){8-9}\cmidrule(lr){10-11}\cmidrule(l){12-13}
& \textit{Avg.}$\uparrow$ & \textit{Borda Avg.}$\uparrow$
& \textit{Avg.}$\uparrow$ & \textit{Borda Avg.}$\uparrow$
& \textit{Avg.}$\uparrow$ & \textit{Borda Avg.}$\uparrow$
& \textit{Avg.}$\uparrow$ & \textit{Borda Avg.}$\uparrow$
& \textit{Avg.}$\uparrow$ & \textit{Borda Avg.}$\uparrow$
& \textit{Avg.}$\uparrow$ & \textit{Borda Avg.}$\uparrow$ \\
\midrule
\multicolumn{13}{l}{\textit{AR (next-token prediction)}} \\
XTTS-v2     & \textbf{0.31} & \textbf{5.14} & \underline{4.18} & \underline{4.70} & 127.84 & 4.89 & \textbf{0.28} & \textbf{5.73} & 4.12 & 4.16 & \textbf{93.15} & \underline{5.50} \\
CosyVoice   & 0.27 & 3.46 & 4.08 & 4.41 & 120.59 & 4.59 & 0.22 & 3.49 & \underline{4.29} & \underline{4.85} & 75.74 & 4.85 \\
CosyVoice 2 & \underline{0.30} & \underline{4.56} & 4.07 & 4.47 & \underline{134.34} & \underline{5.38} & \underline{0.24} & 4.45 & 4.12 & 4.35 & \underline{88.04} & \textbf{5.78} \\
\midrule
\multicolumn{13}{l}{\textit{NAR (flow matching)}} \\
E2 TTS      & 0.27 & 3.26 & 3.40 & 1.87 & 84.91  & 2.11 & 0.20 & 2.78 & 3.33 & 1.57 & 52.35 & 2.18 \\
F5-TTS      & 0.26 & 2.98 & 3.48 & 2.19 & 79.59  & 1.50 & 0.21 & 3.23 & 3.49 & 2.28 & 49.00 & 1.51 \\
ZipVoice    & 0.29 & 4.55 & 3.91 & 3.85 & 114.52 & 3.93 & 0.22 & 3.55 & 3.99 & 4.09 & 58.56 & 2.88 \\
\midrule
\multicolumn{13}{l}{\textit{NAR (masked generative modeling)}} \\
MaskGCT     & 0.28 & 4.04 & \textbf{4.76} & \textbf{6.51} & \textbf{139.75} & \textbf{5.61} & \underline{0.24} & \underline{4.78} & \textbf{5.13} & \textbf{6.70} & 80.36 & 5.30 \\
\bottomrule[1pt]
\end{tabular}
}
\vspace{-1.5em}
\label{tab:benchmark}
\end{table*}

\section{\ours{} Metric}\label{sec:\ours{}}
Given a zero-shot TTS system, we aim to quantify prosody diversity between two generated speech samples conditioned on the same text and reference speech prompt, using distinct random seeds.

\noindent\textbf{Formulation}\quad
Let $\mathbf{X}_1$ and $\mathbf{X}_2$ denote two synthesized speech samples.
To remove the influence of leading and trailing silences, a pre-trained VAD model\footnote{\url{https://github.com/snakers4/silero-vad}} is employed to trim the raw waveforms:
\begin{equation}
\tilde{\mathbf{X}}_1 = \mathbf{X}_1[t_1^\mathrm{start} : t_1^\mathrm{end}], \qquad
\tilde{\mathbf{X}}_2 = \mathbf{X}_2[t_2^\mathrm{start} : t_2^\mathrm{end}],
\end{equation}
where $t_1^\mathrm{start}$, $t_2^\mathrm{start}$, $t_1^\mathrm{end}$, and $t_2^\mathrm{end}$ are the onset and offset timestamps predicted by the VAD model. The resulting $\tilde{\mathbf{X}}_1$ and $\tilde{\mathbf{X}}_2$ are the corresponding silence-trimmed synthesized speech samples.

For speech tokenization, we use a self-supervised speech representation model combined with $k$-means clustering to discretize the silence-trimmed speech into token sequences $\mathbf{c}_1$ and $\mathbf{c}_2$:
\vspace{-0.3em}
\begin{equation}
\mathbf{c}_1 = \mathrm{Encode}_\mathrm{spch}(\tilde{\mathbf{X}}_1), \qquad
\mathbf{c}_2 = \mathrm{Encode}_\mathrm{spch}(\tilde{\mathbf{X}}_2).
\vspace{-0.3em}
\label{eq:tokenization}
\end{equation}

Prosodic variation is quantified as a weighted Levenshtein distance~\cite{levenshteindistance} between discrete speech token sequences:
\vspace{-0.3em}
\begin{equation}
\mathrm{\text{DS-WED}}(\mathbf{c}_1,\mathbf{c}_2) 
= \min_{\pi \in \mathcal{A}(\mathbf{c}_1,\mathbf{c}_2)}
    \sum_{(i,j,o) \in \pi} w_o \, c_o(c_{1,i}, c_{2,j}),
\label{eq:\ours{}}
\vspace{-0.3em}
\end{equation}
where $\pi$ is an edit path in the alignment set $\mathcal{A}$, $o \in \{\mathrm{sub},\mathrm{ins},\mathrm{del}\}$ is the edit operation, $c_o(\cdot)$ denotes the corresponding edit cost, and $w_o$ is an operation-dependent weight, set to $1$ in all experiments.

\vspace{0.1em}
\noindent\textbf{Discussion}\quad
\ours{} is designed to be representation-agnostic. We instantiated it with semantic tokens~\cite{discretespeech} for following reasons:
\begin{itemize}[leftmargin=*, itemsep=0pt, topsep=0pt]
    \item Semantic tokens obtained by $k$-means clustering of SSL representations from HuBERT~\cite{hubert} and WavLM~\cite{wavlm} have been demonstrated to effectively capture prosodic information~\cite{prosodydiscretespeech}.
    \item Supervised semantic tokens from S3Tokenizer~\cite{cosyvoice, cosyvoice2, cosyvoice3} are not considered, since the introduction of sequence-level ASR loss like CTC~\cite{CTC} tends to distort token duration information.
    \item Acoustic tokens from EnCodec~\cite{encodec} are not considered, as they retain low-level signal details that are not relevant to prosody.
    \item Operation weights can be tuned to perceptual sensitivity. Listening tests indicate higher sensitivity to intonation and word stress than to pause duration, suggesting that assigning larger weights to substitution operations may better align with human judgment.
    \item The weighted edit distance can be interpreted as the minimum perceptually prosodic modifications needed to transform one speech into another at the discrete level.
\end{itemize}
\section{Experiments}

\subsection{Correlation Analysis with Human Judgments}
\noindent\textbf{Setup}\quad
We evaluate correlations between human ratings and the three objective metrics \ours{}, log $F_0$ RMSE, and MCD on the ProsodyEval dataset across all groups.
For \ours{}, speech is discretized by applying a 50-cluster $k$-means model trained on LibriSpeech 960h to the hidden embeddings from the 8th Transformer encoder layer.
Since PMOS ratings are relative within groups and not comparable across groups, we compute Pearson correlation for each group and then aggregate the group-wise correlations using Fisher's $Z$ transformation.
Statistical significance is tested with a two-sided one-sample $t$-test against zero, and 95\% confidence intervals are obtained in the Fisher space and back-transformed.

\noindent\textbf{Results}\quad
As shown in Table~\ref{tab:corr}, PMOS exhibits the strongest correlation with \ours{} ($\bar{r}=0.77$), followed by a substantial correlation with MCD ($\bar{r}=0.66$).
The correlation with log $F_0$ RMSE is weaker but still significant ($\bar{r}=0.30$), suggesting that pitch deviations contribute to perceived differences but capture only part of the variability.
Overall, these results demonstrate that \ours{} aligns most closely with subjective human judgments, substantially outperforming widely used acoustic metrics.

\subsection{Efficiency Analysis}
\noindent\textbf{Setup}\quad
Computational efficiency is measured on ProsodyEval by Real-Time Factor (RTF), computed as processing time divided by average speech-pair duration, on NVIDIA A100 with batch size 1.

\noindent\textbf{Results}\quad
Log $F_0$ RMSE reaches an RTF of 0.549, and MCD reaches an RTF of 0.203. Both rely on signal-processing front-ends and DTW alignment, which are CPU-bound and difficult to accelerate on GPUs. In addition, log $F_0$ RMSE requires mel-cepstrum computation for DTW, adding extra overhead.
In contrast, \ours{} involves only a forward pass through a pretrained speech-SSL encoder followed by $k$-means clustering and edit distance at the discrete level, achieving an RTF of 0.110. It is GPU-friendly and can be further accelerated by batching.
Overall, \ours{} is scalable for large-scale evaluation and practical for speech data engineering.

\subsection{Ablation Studies of \ours{}}
\noindent\textbf{Setup}\quad
We study the effect of the SSL backbone, the encoder layer, and the number of $k$-means clusters on the correlation between \ours{} and human ratings, using the ProsodyEval dataset.

\begin{figure}[t]
\vspace{-0.6em}
\centering
\includegraphics[width=\linewidth]{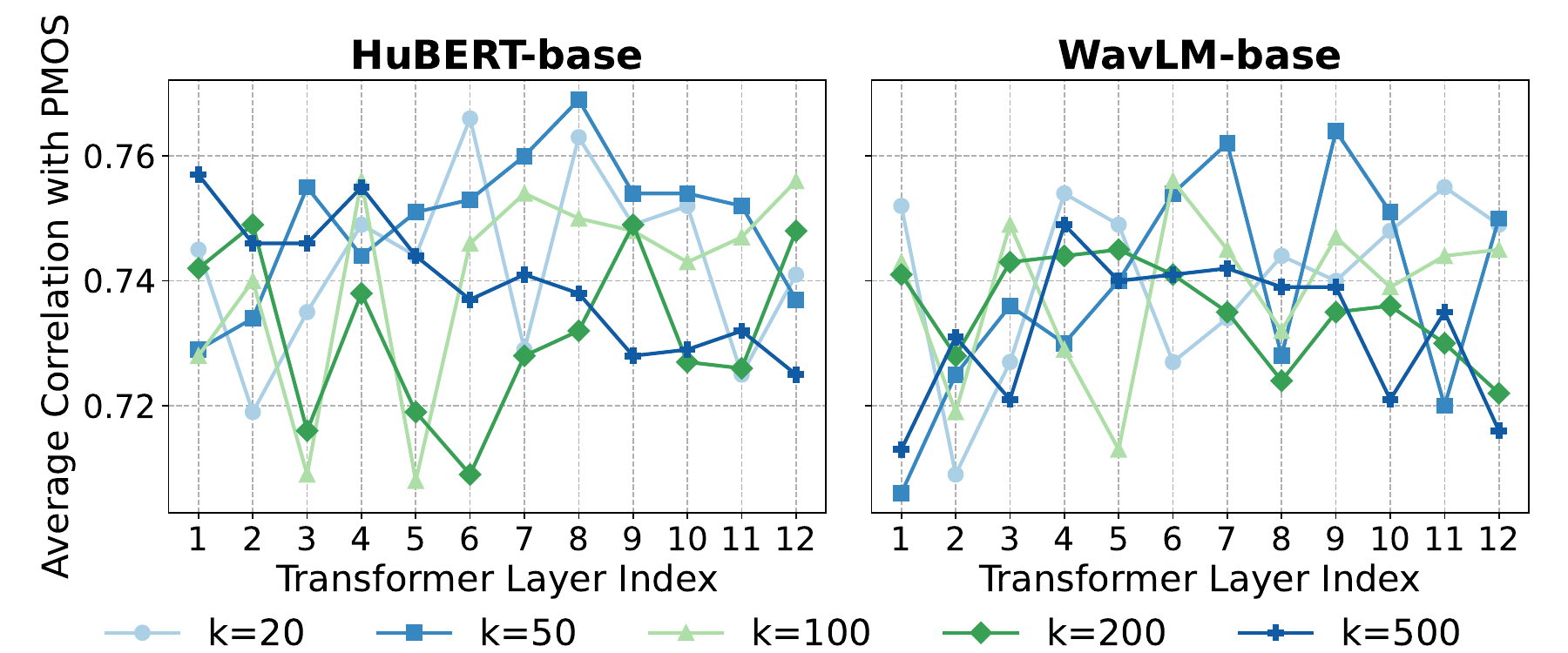}
\vspace{-2.4em}
\caption{
Ablation of average correlations between \ours{} and human ratings across models, layers, and cluster sizes on ProsodyEval.
}
\vspace{-1.5em}
\label{fig:ablation}
\end{figure}

\noindent\textbf{Results}\quad
Figure~\ref{fig:ablation} shows that \ours{} remains quite robust across different layers, models, and vocabulary sizes, with correlations consistently around 0.7.
Middle layers 6-9 achieve stronger correlations, consistent with the richer encoded prosody information. 
Relative smaller cluster sizes perform best, while larger ones reduce peak correlations, making the edit-distance calculation overly sensitive and misaligned with human perceptual sensitivity to prosody.
Overall, WavLM-base provides more stable correlations, while HuBERT-base exhibits larger variance. The 8th layer of HuBERT-base using 50 clusters achieves the strongest correlations.
We also tried the large versions, which yield slightly higher correlations but at much greater cost, hence we use the base versions for all experiments.

\subsection{Benchmarking Zero-shot TTS Systems}
\noindent\textbf{System Details}\quad  
We evaluate the following open-source zero-shot TTS systems across three representative paradigms: next-token prediction, flow matching, and MGM.
\begin{itemize}[leftmargin=*, itemsep=0pt, topsep=0pt]
\item \textbf{CosyVoice}~\cite{cosyvoice} \& \textbf{CosyVoice 2}~\cite{cosyvoice2}: Two-stage AR+flow-matching systems trained on 166.8k hours of multilingual data. We evaluate CosyVoice-300M\footnote{\url{https://www.modelscope.cn/iic/CosyVoice-300M}} and CosyVoice2-0.5B\footnote{\url{https://www.modelscope.cn/iic/CosyVoice2-0.5B}}.

\item \textbf{MaskGCT}~\cite{maskgct}: A two-stage NAR MGM system\footnote{\url{https://huggingface.co/amphion/MaskGCT}} with 695M text-to-semantic and 353M semantic-to-acoustic models, trained on 100k hours of Chinese and English speech from Emilia~\cite{emilia}.

\item \textbf{E2 TTS}~\cite{e2tts} \& \textbf{F5-TTS}~\cite{f5tts}: Fully NAR flow-matching systems with 333M\footnote{\url{https://huggingface.co/SWivid/E2-TTS}} and 336M\footnote{\url{https://huggingface.co/SWivid/F5-TTS}} parameters, trained on Emilia 100k hours.

\item \textbf{ZipVoice}\cite{zipvoice}: A fully NAR flow-matching system\footnote{\url{https://huggingface.co/k2-fsa/ZipVoice}} with 123M parameters, trained on Emilia 100k hours.

\item \textbf{XTTS}~\cite{xtts}: A two-stage AR+VQ-VAE system trained on multilingual data. We evaluate XTTS-v2\footnote{\url{https://huggingface.co/coqui/XTTS-v2}}, which is trained on a similar amount of data as XTTS-v1, around 27k hours.
\end{itemize}

\noindent\textbf{Setup}\quad
Our evaluations are conducted on two widely used benchmarks: professionally read audiobooks LibriSpeech \textit{test-clean} and crowdsourced read speech Seed-TTS \textit{test-en}.
To measure prosody diversity, we report two conventional acoustic metrics, log $F_0$ RMSE and MCD, together with our proposed \ours{}. Each metric is computed in two ways: (1) micro average across all samples, and (2) rank-based Borda aggregation, where systems are ranked within each group and assigned scores from seven for the best system to one for the worst, and the average score across groups is reported, thereby eliminating the influence of absolute metric values.

\noindent\textbf{Results}\quad
As shown in Table~\ref{tab:benchmark}, AR systems consistently outperform NAR flow-matching systems in prosody diversity. However, when compared with NAR MGM systems, AR systems show comparable performance. MaskGCT even surpasses all AR systems on LibriSpeech and remains competitive on Seed-TTS.

\begin{table}[t]
\centering
\small
\caption{
Effect of duration perturbation (DP) on prosody diversity of NAR TTS systems on LibriSpeech \textit{test-clean} and Seed-TTS \textit{test-en}, using \ours{}, computed as micro-average (\textit{Avg.}).
}
\vspace{2pt}
\renewcommand\tabcolsep{5pt}
\resizebox{\linewidth}{!}{
\begin{tabular}{lcc}
\toprule[1pt]
\multirow{2}{*}{\textbf{System}} & \textbf{LibriSpeech \textit{test-clean}} & \textbf{Seed-TTS \textit{test-en}} \\
\cmidrule(l){2-2} \cmidrule(l){3-3}
 & \ours{} \textit{Avg.}$\uparrow$ & \ours{} \textit{Avg.}$\uparrow$ \\
\midrule
F5-TTS & 79.59 & 49.00  \\
F5-TTS w/ DP & 100.88$_{\textcolor{green!80!black}{+26.7\%}}$ & 62.95$_{\textcolor{green!80!black}{+28.5\%}}$ \\
\midrule
MaskGCT & 139.75 & 80.36 \\
MaskGCT w/ DP & 159.10$_{\textcolor{green!80!black}{+13.8\%}}$ & 92.71$_{\textcolor{green!80!black}{+15.4\%}}$ \\
\bottomrule[1pt]
\end{tabular}
}
\vspace{-1.5em}
\label{tab:duration}
\end{table}

\begin{table}[t]
\centering
\small
\caption{
Effect of DPO on prosody diversity of zero-shot TTS systems on LibriSpeech \textit{test-clean} and Seed-TTS \textit{test-en}, using \ours{}, computed as micro-average (\textit{Avg.}).
}
\vspace{2pt}
\renewcommand\tabcolsep{5pt}
\resizebox{\linewidth}{!}{
\begin{tabular}{lcc}
\toprule[1pt]
\multirow{2}{*}{\textbf{System}} & \textbf{LibriSpeech \textit{test-clean}} & \textbf{Seed-TTS \textit{test-en}} \\
\cmidrule(l){2-2} \cmidrule(l){3-3}
 & \ours{} \textit{Avg.}$\uparrow$ & \ours{} \textit{Avg.}$\uparrow$ \\
\midrule
CosyVoice 2 & 134.34 & 88.04 \\
CosyVoice 2 w/ DPO & 109.09$_{\textcolor{red!80!black}{-18.8\%}}$ & 71.64$_{\textcolor{red!80!black}{-18.6\%}}$ \\
\midrule
MaskGCT & 139.75 & 80.36 \\
MaskGCT w/ DPO & 135.75$_{\textcolor{red!80!black}{-2.9\%}}$ & 77.80$_{\textcolor{red!80!black}{-3.2\%}}$ \\
\bottomrule[1pt]
\end{tabular}
}
\vspace{-1.5em}
\label{tab:dpo}
\end{table}
\begin{table}[t]
\centering
\small
\caption{
Average Pearson ($\bar{r}$) correlations between LALM-as-Judges and prosody-related metrics, aggregated across groups via Fisher’s $Z$ transformation.
Values in brackets denote 95\% confidence intervals, computed in the Fisher space and back-transformed.
* marks correlations statistical significance at $p < 0.05$.
}
\vspace{2pt}
\renewcommand\tabcolsep{2pt}
\resizebox{\linewidth}{!}{
\begin{tabular}{lcccc}
\toprule[1pt]
 & PMOS & log $F_0$ RMSE & MCD & \ours{} \\
\midrule
Gemini 2.5 Pro & 0.27*{\tiny [0.16,\;0.38]} & 0.10{\tiny [-0.08,\;0.27]} & 0.16*{\tiny [0.01,\;0.30]} & 0.22*{\tiny [0.05,\;0.38]} \\
\bottomrule[1pt]
\end{tabular}
}
\vspace{-1em}
\label{tab:speechllm}
\end{table}

\subsection{Insights from Benchmarking and Further Exploration}
\noindent\textbf{\textit{Do AR zero-shot TTS systems generate more prosody diversity than NAR systems?}}
AR systems indeed outperform flow-matching models in prosody diversity but offer no advantage over MGM.
\begin{itemize}[leftmargin=*, itemsep=0pt, topsep=0pt]
    \item AR models generate speech sequentially, providing explicit temporal modeling and natural variation in duration.
    In contrast, NAR flow-matching systems such as E2-TTS, F5-TTS, and ZipVoice pursue architectural simplicity by adopting implicit alignment.
    When regression objectives are applied to these weakly aligned and entangled representations, the models collapse toward mean predictions of highly multimodal and diverse prosodic patterns, resulting in blurry and over-smoothed outputs~\cite{prosodytts}.
    Average upsampling-based alignment combined with a text encoder in ZipVoice partly alleviates this issue, but the problem remains non-trivial.
    By comparison, MGM introduces stochasticity through iterative mask-and-prediction.

    \item All NAR systems are trained on the same 100k-hour Emilia corpus, yet MaskGCT performs best in \ours{} while F5-TTS ranks lowest.  
    Moreover, XTTS-v2, trained on less than one-third of the data, still shows superior prosody diversity, suggesting that modeling paradigm rather than data scale is the dominant factor.
\end{itemize}

\vspace{0.5em}
\noindent\textbf{\textit{To what extent does duration variation shape prosody diversity in NAR zero-shot TTS systems?}}
We apply duration perturbation (DP) with factors 0.8, 0.9, 1.0, 1.1, and 1.2 to audio samples within each group.
Prior work~\cite{maskgct, palle} shows duration variations in this range have little effect on intelligibility.
As shown in Table~\ref{tab:duration}, DP consistently increases prosody diversity for two NAR systems, indicating that duration variation during inference is a key factor shaping prosody.
Notably, F5-TTS exhibits a relative increase of nearly 30\% with DP, yet still lags behind AR and MGM systems without DP, suggesting that prosodic monotony in flow-matching systems with implicit alignment stems from inherent architectural limitations.

\vspace{0.5em}
\noindent\textbf{\textit{Does reinforcement learning (RL) reduce prosody diversity in zero-shot TTS systems?}}
We evaluate the prosody diversity of zero-shot TTS systems from~\cite{intp}, which are aligned via Direct Preference Optimization (DPO) on the INTP dataset~\cite{intp} for intelligibility, using vanilla DPO for AR and extended DPO~\cite{intp} for NAR MGM.
As shown in Table~\ref{tab:dpo}, applying DPO consistently reduces prosodic diversity for AR and NAR systems, reflecting the general tendency of RL to prune variation while amplifying reward-aligned behaviors.

\vspace{0.5em}
\noindent\textbf{\textit{Can large audio language models (LALMs) serve as reliable evaluators of prosodic differences?}}
We test Gemini 2.5 Pro~\cite{Gemini25} by prompting it to rate the relative prosodic difference between two samples within a group of five. It listens to all five samples to form a reference range of variation and then assigns a score from 1 to 5.
As shown in Table~\ref{tab:speechllm}, Gemini's scores show a statistically significant but weak correlation with human ratings, while correlations with objective metrics fluctuate with wide confidence intervals.
Combined with high prompt sensitivity, these findings suggest that Gemini 2.5 Pro is not a reliable evaluator of prosodic variation.
\section{Conclusion}
In this work, we propose \ours{}, a reliable objective diversity metric based on weighted edit distance over semantic tokens, together with ProsodyEval, a human-annotated dataset for assessing prosody diversity in zero-shot TTS.
\ours{} outperforms frequently used acoustic metrics in correlating with human judgments and remains robust in speech discretization across models, layers, and cluster sizes, enabling reliable quantitative evaluation of prosodic variation.
Leveraging \ours{}, we systematically benchmark and analyze mainstream zero-shot TTS systems.  
We quantitatively show that (1) AR systems outperform NAR flow-matching models but not MGM, as duration variation strongly shapes prosody diversity; however, NAR lack duration modeling, and flow-matching models with implicit alignment suffer from inherent limitations, (2) RL via DPO improves intelligibility while reducing prosody diversity, and (3) even advanced LALMs like Gemini 2.5 Pro remain unreliable for prosody evaluation.  
One limitation lies in the cross-lingual applicability of \ours{}, validated only on English.
Looking ahead, ProsodyEval and \ours{} fill a gap in zero-shot TTS evaluation and open new avenues for speech data engineering.

\clearpage
\bibliographystyle{IEEEbib}
\bibliography{refs}

\end{document}